\documentclass[notitlepage,twocolumn,pre,tightenlines,nofootinbib,showpacs,superscriptaddress]{revtex4}

\usepackage{amsmath}
\usepackage{amssymb}
\usepackage{graphicx}
\usepackage{color}

\definecolor{blue}{rgb}{0,0,1}
\definecolor{bleu}{rgb}{0,0,0.8}
\definecolor{bleuf}{rgb}{0,0,0.9}
\definecolor{rougef}{rgb}{0.9,0,0}
\definecolor{green}{rgb}{0,0.5,0}
\definecolor{vert}{rgb}{0,0.8,0}
\definecolor{red}{rgb}{1,0,0}
\definecolor{pink}{rgb}{0.9,0.3,0.7}
\definecolor{azur}{rgb}{0,0.5,0.5}
\definecolor{orange}{rgb}{1,0.5,0.2}
\definecolor{brown}{rgb}{0.5,0,0}

\newcommand{\be}{\begin{equation}}
\newcommand{\ee}{\end{equation}}
\newcommand{\ben}{\begin{equation*}}
\newcommand{\een}{\end{equation*}}
\newcommand{\ba}{\begin{eqnarray}}
\newcommand{\ea}{\end{eqnarray}}

\newcommand{\leg}[1]{\textbf{#1}}

\begin{document}
\graphicspath{{./Figures/}}

\title{Shear Modulus and Dilatancy Softening in Granular Packings above Jamming.}
\author{C. Coulais}
\affiliation{SPHYNX/SPEC, CEA-Saclay, URA 2464 CNRS, 91191 Gif-sur-Yvette, France}
\affiliation{Universit\'e Paris-Sud, CNRS, Lab FAST, Bat 502, Campus Universit\'e, Orsay, F-91405, France}
\affiliation{Huygens-Kamerlingh Onnes Lab, Universiteit Leiden, PObox 9504, 2300 RA Leiden, The Netherlands}
\author{A. Seguin}
\affiliation{SPHYNX/SPEC, CEA-Saclay, URA 2464 CNRS, 91191 Gif-sur-Yvette, France}
\affiliation{Universit\'e Paris-Sud, CNRS, Lab FAST, Bat 502, Campus Universit\'e, Orsay, F-91405, France}
\author{O. Dauchot}
\affiliation{EC2M, ESPCI-ParisTech, UMR Gulliver 7083 CNRS, 75005 Paris, France}

\begin{abstract}
We investigate experimentally the mechanical response to shear of a monolayer of
bi-disperse frictional grains across the
jamming transition. We inflate an intruder inside the packing and use photo-elasticity and
tracking techniques to measure the induced shear strain and stresses at the grain scale. 
We quantify experimentally the constitutive relations for
strain amplitudes as low as $10^{-3}$ and for a range of packing 
fractions within $2\%$ variation around the jamming transition.
At the transition strong nonlinear effects set in : both the shear
modulus and the dilatancy shear-soften at small strain until a critical strain
is reached where effective linearity is recovered. 
The scaling of the critical strain and the associated critical stresses on
the distance to jamming are extracted. 
We check that the constitutive laws, together with mechanical equilibrium, correctly predict to the observed stress and strain
profiles. These profiles exhibit a spatial crossover between an effective linear
regime close to the inflater and the truly nonlinear regime away from it. The
crossover length diverges at the jamming transition. 
\end{abstract}


\maketitle


{\em Introduction.} --- 
Understanding the mechanical properties of dense packings of athermal particles,
such as grains, foams and emulsions, remains a conceptual and practical
challenge. When decreasing the packing fraction $\phi$, these intrinsically out-of-equilibrium systems lose their rigidity
at the so-called jamming transition, $\phi=\phi_J$, when the confining pressure approaches zero and 
the particles deformations vanish~\cite{Liu1998,OHern2002,OHern2003,vanHecke2010}. In the case of frictionless
spheres~\cite{OHern2002,OHern2003}, the loss of mechanical stability coincides with the onset of isostaticity : 
the average number of contacts $z$ decreases to its isostatic value, for which the number of geometrical
and mechanical equilibrium constraints exactly match the number of degrees of freedom. 
Approaching the transition, the material becomes more and more
fragile~\cite{Cates1998}, and its linear response, dominated by floppy
modes~\cite{Wyart2005EPL}, exhibits critical
scaling~\cite{OHern2002,OHern2003,Ellenbroek2006,vanHecke2010}. 

In a first step towards the description of such systems, Wyart et al.~\cite{Wyart2005EPL,Wyart2005PRE,Brito2010,Xu2009,During_SoftMatter2013} 
derived a scaling theory of the jamming transition from a marginal stability principle, which 
captures most of its phenomenology.
Recently, marginality has been translated into the adoption of a full replica symmetry
breaking scheme in the formulation of a mean field theory of hard sphere glasses
at high density~\cite{Parisi2010,Berthier2011,Charbonneau2014}. As a result, the theory properly
describes not only the thermodynamic properties of the packing, but also the
structural and dynamical ones, when approaching $\phi_J$. 

The relevance of these theories for real systems remains to be established.
There are very few direct experimental investigations of the scaling regime above jamming.
The average number of contacts has been measured in grains~\cite{Majmudar2007,Coulais2012}, foams~\cite{Katgert2010} and emulsions~\cite{Jorjadze2013} but not with a sufficient accuracy to provide
stringent bounds for the value of the scaling exponent $\delta$.  As for the dynamics and the mechanics, 
rheology below jamming has been studied in vibrated grains~\cite{Dijksman2011}, foams~\cite{Katgert2013} 
and emulsions~\cite{Mansard2011}, but we are not aware of any direct measurements of the elastic moduli dependence 
on the packing fraction when approaching jamming \emph{from above}.

Also, the relevance of the linear response very close to the transition remains a matter of debate~\cite{Schreck2011,Bertrand2014,Goodrich2014}.  At finite shear strain amplitude $\gamma$, non-linear effects become dominant~\cite{Lerner2013,Brito2010,Gomez2012}  and the mechanical response of the system is no
longer relevantly described exclusively by $\Delta z$. 
Finally, while dilatancy effects -- namely the increase of volume or pressure under shear -- are important in sheared 
granular experiments~\cite{Reynolds1885,Bi2011,Ren2013}, they are systematically missed in numerical and theoretical 
studies of soft spheres near jamming.

\begin{figure*}[t!]
\centering
\includegraphics[width=1.6\columnwidth]{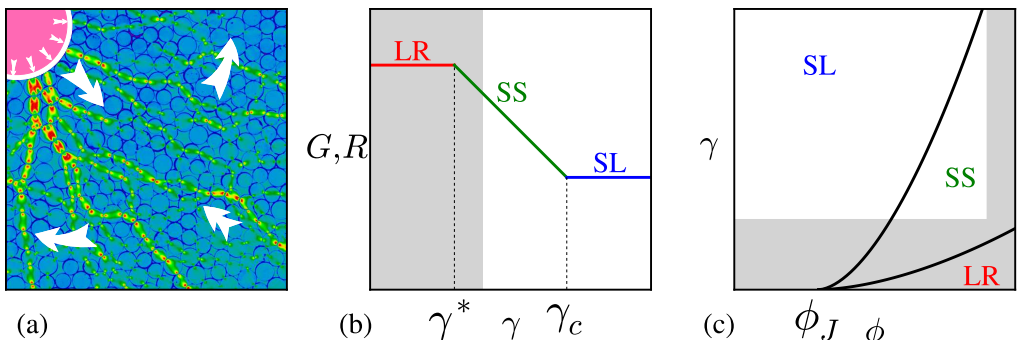}
\vspace{-0.5cm}
\caption{{\bf (a)} Quadrant of combined raw photoelastic and direct light pictures. The intruder (pink) is inflated and induces radial compression and orthoradial stretch (white arrows): the packing is sheared azimuthally. 
{\bf (b)} Sketch of the shear modulus, $G$ and dilatancy coefficient $R$, vs. shear strain $\gamma$.  
In the linear regime (LR, $\gamma<\gamma^*$), not probed here,  both are constant. For $\gamma^*<\gamma<\gamma_c$ both \emph{decrease}, this is a shear softening (SS) regime. For $\gamma>\gamma_c$,  effective linear elasticity (SL) is recovered. 
{\bf(c)} $\phi-\gamma$ parameter space with the different  regimes : both $\gamma^*$ and $\gamma_c$  vanish at Jamming. The gray regions could not be accessed in the present experiment.}
\label{fig:sketches}
\vspace{-0.5cm}
\end{figure*}

In this Letter, taking advantage of the possibility to probe jamming scalings in a weakly vibrated 
monolayer of soft grains~\cite{Ikeda2013,Coulais2012,Coulais2014} -- a notoriously difficult task in thermally agitated colloids~\cite{Ikeda2013,Basu2014} --, we provide the first experimental measurement 
of the elastic response of a 2D packing of grains across the jamming transition.
To do so we apply an inhomogeneous shear by inflating an intruder in the center of a monolayer
of bi-disperse frictional grains (fig.~\ref{fig:sketches}a).  We obtain the force network and grain displacements
from photo-elasticity measurements and tracking techniques, and calculate
the stress and strain tensors at the grain scale. The constitutive laws, obtained from a parametric plot of the invariants of the
stress tensor with respect to the shear strain, reveal that linear elasticity
does not apply. Dilatancy is crucial and, above jamming, shear softening occurs at moderate 
strain (fig.~\ref{fig:sketches}b).
Elasticity is effectively recovered only for strains larger than a critical strain, which scales
with the distance to jamming and eventually vanishes at $\phi_J$ (fig.~\ref{fig:sketches}c). 
We compute the strain profiles from the inferred constitutive laws and show that
they match the experimental profiles and display a spatial
crossover between the two regimes. The crossover length diverges like
$\Delta\phi^{-0.85}$ when the system (un)jams.

{\em Setup and Protocol.} --- The setup is adapted from~\cite{Coulais2012,Coulais2014}.
A bi-disperse layer of $8166$ photo-elastic disks of diameter $4$ and
$5$~mm is confined in a rectangular frame. A wall piston allows to precisely tune
the packing fraction $\phi$. The grains lie on
a glass plate which can be vibrated with an amplitude of $1$~cm at a frequency
of $10$~Hz perpendicularly to the direction of the wall piston.
The inflater is made of a brass spacer, equipped with $9$ radial
pistons, is surrounded by an O-ring of diameter $2 r_I = 26.3$~mm and connected
to a pressure switch. When the pressure is increased inside the spacer,
the pistons push the O-ring radially, ensuring a uniform radial dilation, up to
$2(r_I+a) = 28.5$~mm. When the pressure is switched off, the elasticity of the
O-ring brings back the inflater to its initial diameter. The dilation rate
$a^*=a/ r_I \in [1-10]\%$.

Varying both the strain amplitude and packing fraction, we record the stress response
following a precise protocol. First we introduce the inflater at
the center of the packing at low packing fractions. We then compress the packing 
into a highly jammed state while vibrating the bottom plate
(see~\cite{Coulais2014} for details). We stop the vibration and start
acquiring images while increasing the size of the intruder using steps of $1.5\%$. At
the end, we let the inflater recover its initial size, turn on the vibration,
stepwise decrease the packing fraction and start the next measurement loop. 
The vibration steps homogenize the stresses between change of packing fraction, 
while keeping the packing structure identical~\cite{Coulais2012,Coulais2014}.

The photo-elastic grains are backlit with a large, uniform,
circularly polarized light source. Pictures are taken using a high-resolution 
CCD camera. We record both photo-elastic and position information by 
alternating between cross-polarized and direct pictures using a cross polarizer mounted on 
a synchronized step motor (see~\cite{Coulais2014} for
details). We process these images with standard segmentation, tracking and tessellation
techniques, to obtain the displacement field and the force
network~\cite{Coulais2014}. We then compute the strain tensor $\boldsymbol{\epsilon}$ and the
stress tensor $\boldsymbol{\sigma}$ fields at the grain scale~\cite{suppmat,Drescher1972,Cundall1982,Cambou2000,Bi2011}.
Having checked that these tensors share the same eigenvectors~\cite{Cortet2009}, 
we restrict the analysis to their first and second invariants : 
the dilatation $\varepsilon = \frac{1}{2} \sum\limits_{k}  \epsilon_{kk} $,
the pressure $P =- \frac{1}{2} \sum\limits_{k} \sigma_{kk}$, the shear strain
$\gamma=\sqrt{\frac{3}{2}\sum\limits_{i,j} \left(\epsilon_{ij}-\varepsilon\delta_{ij}\right)^2}$
and the shear stress $\tau=\sqrt{\frac{3}{2}\sum\limits_{i,j}\left(\sigma_{ij}+P\delta_{ij}\right)^2}$ where
$\delta_{ij}$ is the Kronecker symbol. In the following, $P$ and $\tau$ are normalized by the contact stiffness 
$k=1$~N/mm and the length unit is the diameter of the small grains $s=4$~mm. The stress and strain tensors
are respectively measured with a resolution of $10^{-4}$ and $10^{-3}$.

\begin{figure}[t!] 
\center
\includegraphics[width=\columnwidth]{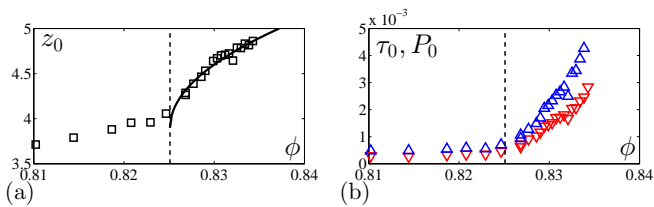}
\vspace{-0.5cm}
\caption{{\bf Initial stress state.} (color online)  \leg{(a)} Initial average contact number $z_0$  ($\square$); \leg{(b)}, pressure $P_0$ ($\triangle$) and shear stress $\tau_0$ ($\triangledown$) vs. $\phi$. 
The solid line is a fit to $z_0=z_p(\phi-\phi_J)^{0.5}+z_J$, with $\phi_J=0.8251\pm0.0009$, $z_p=10.0\pm 0.5$, 
and $z_J=3.9\pm 0.1$. The dashed line indicates $\phi_J$.}
\vspace{-0.5cm}
\label{fig:initialstress}
\end{figure}

{\em Initial state.} --- For each packing fraction, before inflating the
intruder, the system is characterized by an initial state, with force chains spanning the whole system.
This compressed state above jamming, which has been studied in detail
before~\cite{Coulais2014}, is statistically homogeneous. The average contact number $z_0$ 
is essentially constant at low packing fraction (see fig.~\ref{fig:initialstress}a). 
At intermediate packing fraction, it exhibits a kink from where 
it increases sub linearly. We identify the location of the kink with the jamming transition at packing fraction
$\phi_J=0.8251\pm 0.0009$. One should not be surprised to observe a finite $z_0$ below
jamming : when the vibration is turned off, the structure is quenched abruptly
from a vibrational state where the averaged number of contact need not be zero.
The sub-linear increase of $z_0$ with packing fraction is compatible with the
one obtained in simulations of frictional particles~\cite{Katgert2010,Somfai2007}. 
The initial pressure $P_0$ also increases above jamming from a small residual
value below jamming, again inherited from the vibrational state (see fig.~\ref{fig:initialstress}b).
Since the packing is compressed by moving only one lateral wall, the compression is not isotropic. 
The packing conserves some anisotropy clearly evidenced by the existence of a
residual shear stress $\tau_0$ proportional to the pressure $P_0$ (see fig.~\ref{fig:initialstress}b).
However the ratio $\tau_0/P_0$ remains smaller than one, as expected for packings where compressive
stresses dominate. 
An important feature of the present geometry is that the azimuthally invariant mechanical driving 
integrates out the anisotropic fluctuations~\cite{Schroder2010,Goodrich_arxiv2014}.

\begin{figure}[t!] 
\center
\includegraphics[width=\columnwidth]{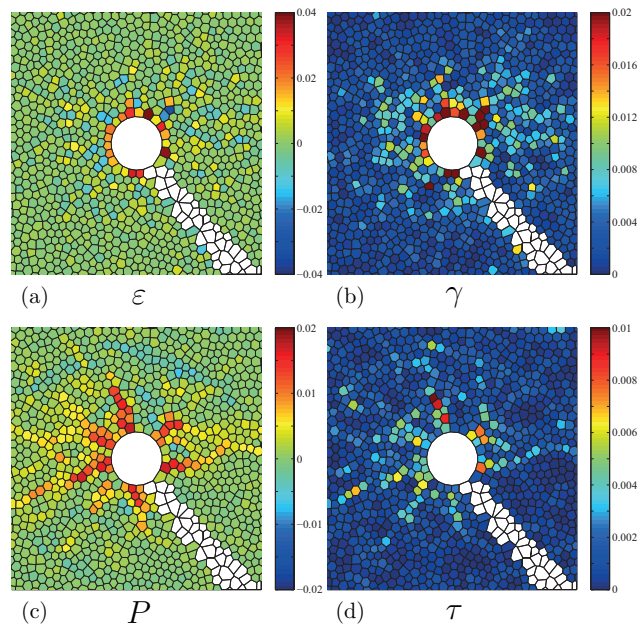}
\vspace{-0.5cm}
\caption{{\bf Maps of the strain and stress invariants.} (color online) Maps of
dilation, $\varepsilon$ ,\leg{(a)}, shear strain, $\gamma$, \leg{(b)}, pressure,
$P$, \leg{(c)} and shear stress, $\tau$, \leg{(d)}, for $\phi=0.8294$ and
$a^*=4.4\times10^{-2}$. The uncolored grains sit below the pneumatic
tube connected to the intruder, which masks the field of view.}
\vspace{-0.5cm}
\label{fig:Maps}
\end{figure}

{\em Response to inflation.} ---  Henceforth, we consider the excess of
pressure $P$ and shear stress $\tau$ produced while inflating the intruder,
namely the difference between the stress measured at the initial state and those measured at each 
$a^*$.  Assuming linear elasticity, $P =- K  \varepsilon$ and $\tau = 2G
\gamma$ (where $K$ is the bulk modulus and $G$ the shear modulus),
the inflation of a disk, in an unconfined geometry induces an azimuthally
invariant shear, which decreases radially with the distance $r$ from the center
of the intruder $ \tau \sim G \gamma \sim a^* /r^2$. 
Figure~\ref{fig:Maps} displays the four maps of the two strain (top row) and two
stress (bottom row) invariants for a typical packing fraction above jamming and
a typical $a^*$ $(4.4\times 10^{-2})$. 
Apart from the spatial fluctuations inherent to the local response of a
disordered material, one observes that the axisymmetry of the loading is
conserved in the response. Furthermore, the response intensity decreases with the
distance from the intruder and we could observe no sign of the lateral walls. In
other words, the hypothesis of an infinite cell is rather well verified (note
that the images shown here represent only one third of the length of the whole
sample). 
Close to the intruder a significant \emph{dilation} occurs because of the boundary
condition geometrical mismatch; but the rest of the packing compresses slightly
and ensures the conservation of the overall volume: the dilation $\varepsilon$ fluctuates
around $6\times 10^{-5}$ with a standard deviation $3\times 10^{-3}$ (fig.~\ref{fig:Maps}a): 
the material is essentially incompressible.  From now on, we shall remove the first shell
around the intruder from the analysis and assume incompressibility, that is $\varepsilon = 0$. 
The second significant observation is that the pressure deviates significantly
from the elastic response : there are regions of intense pressure, which do not
correspond to any sort of intense compression. This pressure field is thus
induced by the shear; it is a manifestation of dilatancy for experiments
conducted at constant volume, a well known effect in granular media~\cite{Reynolds1885}. 
The dilatancy coefficient at constant pressure is related to that at constant volume
by the bulk modulus~\cite{Tighe_2014}.
Finally, whereas the spatially averaged pressure varies linearly with $a^*$,
the spatially averaged shear strain increases faster than $a^*$. This is a first
indication of the nonlinear nature of the material. We checked however that the
shear work $\tau \gamma$ averaged over space scales with ${a^*}^2$.  
The above observations were qualitatively similar for all packing fractions.

\begin{figure}[t!] 
\center
\includegraphics[width=\columnwidth]{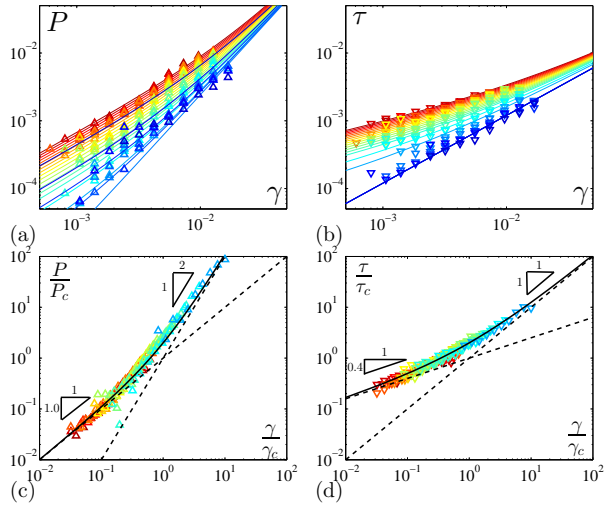}
\vspace{-0.5cm}
\caption{{\bf Constitutive laws.} (color online)
Pressure, $P$ \leg{(a)}, and shear stress,
$\tau$ \leg{(b)},  vs. shear strain, $\gamma$, for 21 packing fractions $\phi
\in [0.8102-0.8343]$. The solid lines are given by Eqs.~(\ref{eq:law1}-\ref{eq:law2}). 
Color code spans from blue to red with increasing packing fractions. 
\leg{(c)} and \leg{(d)}: same data as \leg{(a)} and \leg{(b)} rescaled by $\gamma_c(\phi), P_c(\phi)$ and
$\tau_c(\phi)$. The solid lines are given by the rescaled version of Eqs.~(\ref{eq:law1}-\ref{eq:law2}) and the dashed lines indicate  the asymptotic regimes.}
\vspace{-0.5cm}
\label{fig:Jamming_cross}
\end{figure}

{\em Constitutive laws.} --- 
We now come to the quantitative analysis of the constitutive laws $\tau(\gamma,\phi)$  and $P(\gamma,\phi)$. We collect all data points $P(r,\theta)$ and $\tau(r,\theta)$ vs. $\gamma(r,\theta)$ --- $(r,\theta)$ are the polar coordinates --- into averages corresponding to binned values of $\gamma$. Fig.~\ref{fig:Jamming_cross}a and fig.~\ref{fig:Jamming_cross}b display the obtained shear stress $\tau$ and pressure $P$ versus the shear strain $\gamma$ for different packing fractions. Below jamming, both the shear stress $\tau$ and the pressure $P$ exhibit the simple expected dependence on the shear strain: $\tau = 2 G_0 \gamma$, and $P=R_0 \gamma^2$. Above jamming nonlinearities take place in the form of a significant shear softening of both the shear modulus and the dilatancy. We find that the best description of the data is given by 
\ba
 P & = & \left[R_0+R_{nl}(\Delta\phi,\gamma)\right]\, \gamma^2 \label{eq:law1}\\
 \tau & = &2 \left[G_0+G_{nl}(\Delta\phi,\gamma)\right]\, \gamma \label{eq:law2}
\ea  
\noindent with $\Delta\phi = \phi - \phi_J$, $G_0=6.0\pm0.2\times10^{-2}$, $R_0=1.2\pm0.1\times 10^1$ and
\ba
R_{nl}(\Delta\phi,\gamma) & = & \left\lbrace\begin{array}{c}
                              0 \quad\quad\quad\quad\quad  {\rm for}\, \phi < \phi_J \nonumber \\
                              a \Delta\phi^\mu \gamma^{\alpha-2} \quad {\rm for}\, \phi > \phi_J
                            \end{array} \right. ,
                            \\
G_{nl}(\Delta\phi,\gamma) & = & \left\lbrace\begin{array}{c}
			       0 \quad\quad\quad\quad\quad {\rm for}\, \phi < \phi_J \nonumber \\
			        b \Delta\phi^\nu \gamma^{\beta-1}\quad  {\rm for}\, \phi > \phi_J
                                \end{array} \right. ,
\ea
\noindent
with $\mu=1.7\pm 0.1,\, \alpha=1.0\pm 0.1,\, a= 8.1\pm 0.3\times10^{-2},\, \nu=1.0\pm 0.1,\, \beta=0.4\pm 0.1,\,
b=7.5\pm 0.3\times10^{-1}$. From the above relations, one obtains the rescaling shown in
figures~\ref{fig:Jamming_cross}c and \ref{fig:Jamming_cross}d with $\gamma_c \sim \Delta\phi^\zeta$,
$\tau_c=2G_0 \gamma_c$ and $P_c=R_0 \gamma_c^2$. Despite the fact that the exponent 
pairs $(\mu,\alpha)$ and $(\nu,\beta)$ have been obtained
independently, we find that $\zeta = \mu/(2-\alpha)$ and $\zeta =
\nu/(1-\beta)$ lead to the same value $\zeta = 1.7$, as it should be. The above
equations and the related scaling are the key results of the present study. To
our knowledge, this is the first time that non linear elasticity is quantified
precisely approaching the jamming transition of a granular packing. Note that
the "linear" regime observed here should not be confused with the linear
response and should rather be seen as a saturation of the nonlinearities. For
very small strain, $(\gamma \simeq 10^{-6})$, such as those probed in numerical
studies~\cite{OHern2003,DagoisBohy2012}, and much smaller than the lowest 
strain probed here $(\gamma \simeq 10^{-3})$, one expects to recover a linear 
response for all $\Delta\phi>0$~\cite{Goodrich2014}. For strains of experimental relevance,
very  recent numerical studies have reported a crossover from the linear response
at small strains to a shear softening regime, with a exponent 
$\beta\simeq 0.5$ ~\cite{Otsuki2014,TighePrivComm2014}, compatible with the present results. 

\begin{figure}[t!] 
\center
\includegraphics[width=\columnwidth]{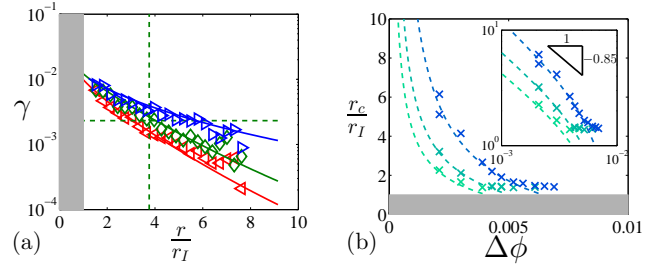}
 \vspace{-0.5cm}
\caption{{\bf Shear strain profiles} (color online)
\leg{(a):} Shear strain profile for ($\triangleright$) ($\phi=0.8208; a^*=0.0374$), ($\diamond$)
($\phi=0.8268;a^*=0.0314$) and ($\triangleleft$) ($\phi=0.8338;a^*=0.0306$). 
The symbols are experimental data and the solid lines come from the integration of eq.(\ref{eq:profile}). 
The green dashed line indicates the crossover for the case ($\phi=0.8268;a^*=0.0314$) 
\leg{(b):} Spatial crossover $r_c(\phi, a^*)/r_I$ (for $a^*=0.0208$ (green), $0.0440$ (turquoise) and 
$0.0681$ (blue) extracted from the experimental profiles ($\times$) in fig~.a and obtained numerically 
from eq.~(\ref{eq:profile}) (dashed lines). \leg{(Inset):} same in log-log axis with the predicted scaling $r_c\sim \Delta\phi^{-0.85}$.  In both figures, the gray zone is the region occupied by the inflater.}
\vspace{-0.7cm}
\label{fig:summary}
\end{figure}

\vspace{0.5cm}
{\em Shear strain profiles.} --- 
We finally proceed to a self-consistency check by integrating the condition of
mechanical equilibrium $\boldsymbol{\nabla}\boldsymbol{\cdot}\boldsymbol{\sigma}
=\boldsymbol{0}$, with the above constitutive laws to derive the expected shear
strain profiles and compare them with those obtained experimentally. We
introduce here the reduced shear strain $\tilde{\gamma}=\gamma/\gamma_c$.
Axisymmetry ensures that $\boldsymbol{\sigma}$ is diagonal in polar coordinate
and independent of the azimuthal coordinate $\theta$.
$\boldsymbol{\nabla}\boldsymbol{\cdot}\boldsymbol{\sigma} =\boldsymbol{0}$ thus
reads: 
\begin{equation}
\frac{P_c (\alpha\tilde{\gamma}^{\alpha-1} + 2\tilde\gamma) + \tau_c
(\beta\tilde{\gamma}^{\beta-1}+1)}{\tilde{\gamma}^{\beta}+\tilde{\gamma}} 
d\tilde{\gamma} = - 2\tau_c\frac{dr}{r}
\label{eq:profile}
\end{equation}
We numerically integrate Eq.~\ref{eq:profile} with the boundary condition
$\tilde\gamma(r=r_I)=a^*/\gamma_c$ and we obtain the profiles plotted in 
figure~\ref{fig:summary}a, together with the experimental data. 
The agreement is excellent, given the absence of any adjustable parameter and the fact that 
we have neglected the confinement at
large $r$. For intermediate values of $\Delta\phi$ and $a^*$, the crossover of
the constitutive law translates into a spatial crossover with a characteristic
length $r_c$ between the saturated linear regime for $r<r_c$, close to the
inflater, and the truly non linear regime for $r>r_c$. An estimate of $r_c$ can
be derived by integrating the above equation in the saturated linear regime and
selecting $\gamma=\gamma_c\;(\tilde{\gamma}=1)$ :
\begin{equation}
\frac{r_c}{r_I} = {\left(\frac{a^*}{\gamma_c}\right)}^{1/2}
\exp\left[\frac{R_0}{2G_0} a^* \left(1-
\frac{\gamma_c}{a^*}\right)\right].
\label{eq:crossover}
\end{equation}
In the limit, $\gamma_c\to 0$, approaching jamming, $r_c \sim \gamma_c^{-1/2} \sim \Delta\phi^{-0.85}$. One can indeed
observe the emergence of this singular behavior on figure~\ref{fig:summary}b, together with the exponential regularization at large $\Delta\phi$. 

{\em Summary-Discussion.} --- We have provided a quantitative characterization of
the elastic response of a 2D packing of grains to the local inflation of an intruder close to jamming.
This specific geometry probes the response to an inhomogeneous shear at
constant volume. Our results highlight the effect of dilatancy and unveil a
nonlinear regime above jamming where both the shear modulus and the dilatancy
coefficient soften. The importance of shear dilatancy in marginal solids was
recently emphasized in~\cite{Tighe_2014}, where it was shown that the Reynolds
coefficient at constant volume $R_V$ scales like $\Delta\phi^{-1/2}$. Here we also
observe a singular behavior, albeit of a different kind since the present
experiment probes the nonlinear softening of the dilatancy. In a different
context, Ren et al.~\cite{Ren2013} report a steep increase of dilatancy under
homogeneous shear as the density of an unjammed packing of grains is increased.
The dilatancy coefficient $R_0$ reported here is very large ($R_0\sim 10^4$~N/m)
and could be seen as a saturation of the divergence reported in~\cite{Ren2013}. 

Finally, the present study uncovers a length scale, $r_c$, which separates the
nonlinear regime from the saturated linear one. Its scaling with the distance to
jamming does not match any scaling reported before for length scales of linear
origin, such as $\ell^*$ or $\ell_c$~\cite{vanHecke2010,During_SoftMatter2013}.
This suggests that $r_c$ could encompass crucial information about the density
of the low energy non-linear excitations reported recently for sphere packings~\cite{Lerner2013}.
Further insights in this matter could come from simulations of point-like
response of the kind reported in~\cite{Ellenbroek2006} albeit in the non linear
regime.


{\em Acknowledgements.} --- We thank B. Tighe, W. Ellenbroek and M.
van Hecke  for discussions. We are grateful to V. Padilla and C. Wiertel-Gasquet for their
skillful technical assistance. This work is supported by the ANR project STABINGRAM 
No. 2010-BLAN-0927-01 and RTRA Triangle de la Physique projects REMIGS2D and COMIGS2D.


\end{document}